# Effects of the DNA state fluctuation on single-cell dynamics of self-regulating gene

Yurie Okabe[1], Yuu Yagi[2], and Masaki Sasai[1,3]

[1]Department of Computational Science and Engineering, and [2]Department of Complex Systems Science, Nagoya University, Nagoya 464-8603, Japan
[3]CREST-Japan Science and Technology Agency, 464-8603, Japan

**Abstract**
A dynamical mean-field theory is developed to analyze stochastic single-cell dynamics of gene expression. By explicitly taking account of nonequilibrium and nonadiabatic features of the DNA state fluctuation, two-time correlation functions and response functions of single-cell dynamics are derived. The method is applied to a self-regulating gene to predict a rich variety of dynamical phenomena such as anomalous increase of relaxation time and oscillatory decay of correlations. Effective "temperature" defined as the ratio of the correlation to the response in the protein number is small when the DNA state change is frequent, while it grows large when the DNA state change is infrequent, indicating the strong enhancement of noise in the latter case.

## I. INTRODUCTION

Dynamical heterogeneity is a prominent feature of chemical reactions in cell. In gene expression of prokaryotic cell, for instance, transcription is suppressed when the promoter site of DNA is occupied by a repressor, while the transcription is enhanced when the repressor is dissociated from DNA. Dynamics of the DNA state can therefore be modeled by the transition between "on" and "off" states. The number of protein molecules in a cell, on the other hand, ranges from $10^0$ to $10^3$, and hence stochastic dynamics of protein synthesis and degradation can be described as the Brownian movement along one-dimensional coordinate representing the protein number. When these two different dynamical fluctuations are coupled to each other, a variety of different phenomena are expected to emerge depending on the precise way of coupling. Sasai and Wolynes[1] pointed out that this coupling resembles interactions between a spin and surrounding atoms in condensed matter, and in such spin-boson problems of condensed matter, difference or similarity in speed of change of constituents has significant effects on dynamics of the whole system. Also in problems of gene expression, difference or similarity in rates of heterogeneous chemical processes should have important influence on the whole cell behavior.

In most of hitherto developed theoretical treatments, difficulty of heterogeneous dynamics in gene expression has been dealt with by assuming that the rate of the DNA state change is fast enough. In such approximate treatments, the fluctuating DNA state has been replaced with the equilibriated state by neglecting the explicit dynamics of DNA state alteration[2-4]. Borrowing the wording from condensed matter physics, this treatment should be called the "adiabatic" approximation. Actual cells, however, may not be in this strong adiabatic limit[5] but resides in weakly adiabatic or nonadiabatic cases. In eukaryotes the DNA state alters much more slowly than in prokaryotes, so that nonadiabaticity is even more important.

As theoretical analyses have shown that the switching rate[6,7] and the oscillatory performance[8] of gene circuits are decisively affected by the nonadiabatic effects, explicit nonadiabatic fluctuation of the DNA state should profoundly affect dynamics of gene expression. Dynamics of gene expression has been experimentally analyzed by monitoring the protein level[2,9], measuring the two-time correlation of the protein number fluctuation[10,11] and by the single-molecule measurement of expressed protein[12] and mRNA[5]. With such progress in experiment, it is now becoming possible to examine effects of nonadiabaticity on single-cell dynamics. It is thus necessary to develop a systematic theoretical method which is comparable with experiments.

Two-time correlation functions of the following type are statistical dynamical quantities which can be examined both in theory and experiment;

$$C^{\mu\nu}(t,t') = \left\langle A^{\mu}(t)A^{\nu}(t')\right\rangle - \left\langle A^{\mu}(t)\right\rangle\left\langle A^{\nu}(t')\right\rangle, \qquad (1)$$

where $A^{\mu}(t)$ can be the number of protein molecules, the number of mRNA molecules, or the DNA state at time $t$ in each cell, and <…> is average over cell lineages in experiment or trajectories in simulation. Quantities which are closely related to $C$ are response functions,

$$R^{\mu\nu}(t,t') = \theta(t-t')\frac{1}{\Delta t}\frac{\delta\left\langle A^{\mu}(t)\right\rangle}{\delta H_{\nu}(t')}\bigg|_{H=0}, \qquad (2)$$

where $\theta(t-t')$ is a step function of $\theta(t-t') = 1$ for $t \geq t'$ and $\theta(t-t') = 0$ for $t < t'$. $R^{\mu\nu}(t,t')$ represents response of the system at time $t$ to a pulse-like perturbation denoted by $H_{\nu}$ which was added to the system during a short period between $t'$ and $t'+\Delta t$. $H_{\nu}$ can be a modulation of the protein synthesis rate or other modulations in rate constants as will be discussed later. When Eq.1 or Eq.2 is applied to the experimental data, <…> should include both the average over the intrinsic fluctuations coming from the smallness of numbers of molecules involved in gene circuits and the average over the extrinsic fluctuations coming from cell growth and division[10,13]. In this paper we focus only on the intrinsic fluctuations, which should have different relaxation time from the extrinsic fluctuations and are hence separable from the contribution of extrinsic fluctuations in the data of time sequence[10]. We expect that the characteristic dynamical features of the gene circuit are reflected in the quantitative functional forms of $C$ and $R$.

$C$ and $R$ or other related quantities have been calculated for molecular systems exhibiting the linear dynamics[14]. In gene circuit, however, dynamics is intrinsically nonlinear because synthesized proteins work as repressors or transcriptional factors to affect the DNA state, and the DNA state in turn determines the rate of protein synthesis. Thus, a transparent theoretical method is necessary to be developed to clarify the effects of nonadiabaticity in nonlinear single-cell dynamics of gene expression. As a first step toward this goal, we take a circuit of self-regulating gene as a simplest example[15].

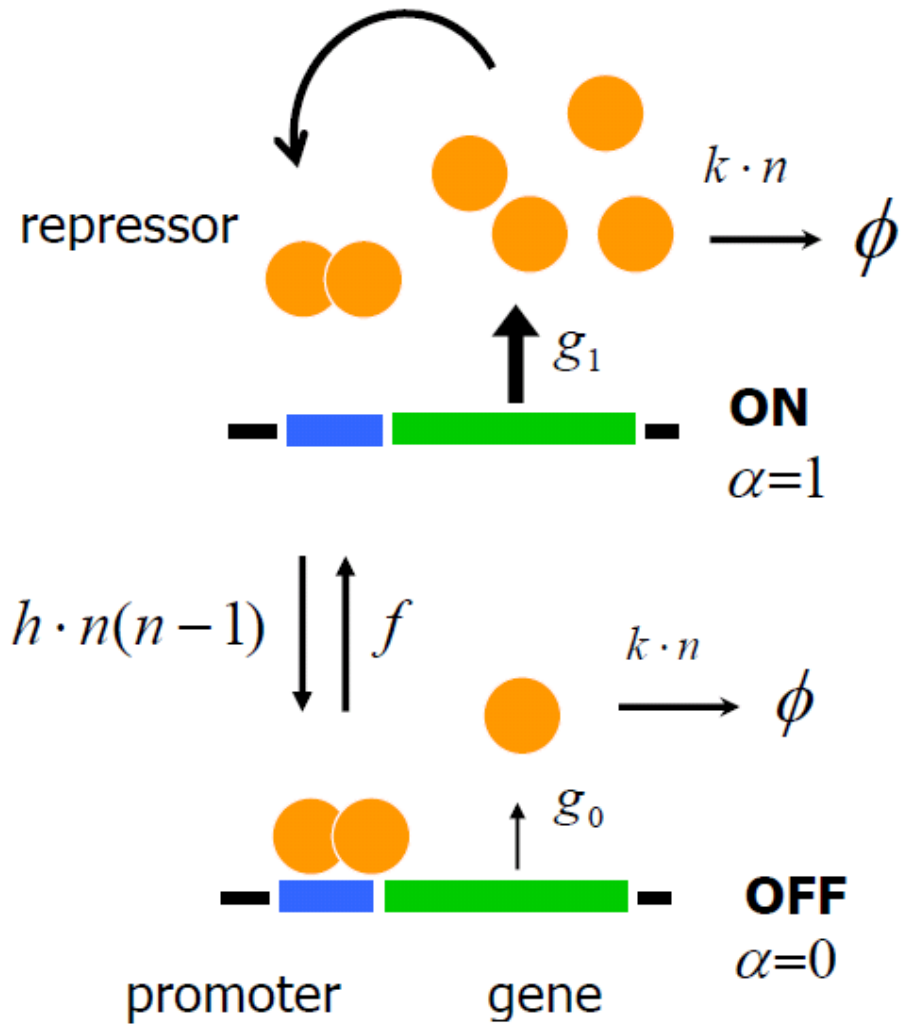

Figure 1
Model of a self-regulating gene. Dimer of the product protein works as a repressor, constituting a negative feedback loop.

## II. MODEL

A model of the gene circuit is shown in Fig.1. A dimer of expressed protein binds as a repressor to the promoter region with the rate $hn(n\text{-}1)$ and is dissociated from DNA with the rate $f$, where $n$ is the number of protein molecules dissociated from DNA. We refer to the DNA state suppressed by repressor as the off state ($\alpha = 0$) and the DNA state without the bound repressor as the on state ($\alpha = 1$). As in previous theoretical treatment[1,3,16-17], degrees of freedom of mRNA are neglected and transcription and translation are described as one step. We assume that $N_b$ molecules of protein are synthesized in a burst with the rate $g_\alpha$ with $g_1 > g_0$. Synthesized protein molecules are degraded with the rate $kn$. Then, the master equation to describe kinetics of Fig.1 is

$$\frac{\partial}{\partial t}P(n,t) = \begin{pmatrix} g_1 & 0 \\ 0 & g_0 \end{pmatrix}\{P(n-N_b,t) - P(n,t)\} - k\{nP(n,t) - (n+1)P(n+1,t)\}$$

$$-\begin{pmatrix} hn(n-1) & 0 \\ 0 & f \end{pmatrix}P(n,t) + \begin{pmatrix} 0 & 0 \\ h(n+2)(n+1) & 0 \end{pmatrix}P(n+2,t) + \begin{pmatrix} 0 & f \\ 0 & 0 \end{pmatrix}P(n-2,t)\,, \tag{3}$$

where $P(n,t) = (P_1(n,t), P_0(n,t))^t$, and $P_\alpha(n,t)$ is the probability that the number of protein molecules dissociated from DNA is $n$ at time $t$ when the DNA state is $\alpha$.

Instead of bear parameters used in Eq.3, we hereafter use the normalized ones, $\omega^{ad} = f/k$, $X^{ad} = (g_1+g_0)/2k$, $\delta X = (g_1-g_0)/2k$, and $X^{eq} = f/h$. $\omega^{ad}$ is a normalized rate of the DNA state alteration, which measures adiabaticity of the present gene circuit: When $\omega^{ad} >>1$, the DNA state alteration is much more frequent than the protein number alteration, which corresponds to the strong adiabatic case. The conventional theories of Langevin dynamics hold only for this strong adiabatic regime[3,4] but we will show in the below that there are a rich dynamical phenomena in the weakly adiabatic regime of $\omega^{ad} \approx 1$ and in the nonadiabatic regime of $\omega^{ad} < 1$. $X^{eq}$ determines the probability that the DNA state is on. $\delta X$ represents the leakage rate of protein synthesis in the off state. $X^{ad}$ is a normalized measure of the rate of protein synthesis, and $N_b X^{ad}$ is a typical number of protein molecules in a cell. To examine the nonadiabatic effects of the DNA state fluctuation, $C$ and $R$ are calculated in wide ranges of $\omega^{ad}$ and $X^{ad}$. For other parameters, we use $X^{eq} = 10^2$, which was estimated from the experimental data on $\lambda$ repressor[18], and $N_b = 10$ as was estimated in a single-molecule measurment[12]. We assume $\delta X = 0.998 X^{ad}$, which corresponds to $g_0/g_1 = 10^{-3}$, allowing small probability of leakage in protein synthesis in the DNA off state.

In the adiabatic limit of $\omega^{ad} >> 1$, the DNA state undergoes on-off transitions so frequently that only the averaged probability of the on or off state does matter. Since the averaged probability can smoothly respond to the change in protein number, the gene circuit of the present model should behave smoothly and stably. In this stable response, both the noise intensity and the relaxation time should be reduced by the negative feedback regulation in the circuit[2, 15]. We may refer to this stable behavior as *normal adiabatic behavior*. When adiabaticity is not strong, however, different behaviors are expected. In the case $\omega^{ad}$ is small and $X^{ad}$ is large, the incidental transition of the DNA state from the off to the on state brings about a burst production of proteins, which should lead to the large fluctuation in the protein number. Hence, the large enhancement of noise is expected in the nonadiabatic regime. We may refer to this noise enhancement as *static nonadiabatic anomaly*. When $\omega^{ad}$ is small and $X^{ad}$ is not so large, on the other hand, the protein production in the DNA on state is not extremely active, allowing the intermittent on-off fluctuation of the DNA state. This intermittent fluctuation should cause the long-time relaxation in fluctuation and response of the protein number. We may refer to this anomaly as *dynamical nonadiabatic anomaly*.

In the following part of the paper, we develop a dynamical mean-field theory to confirm the existence of expected static and dynamical anomalies in this model and to answer questions on parameter regions these anomalies take place and on whether there is a symptom of these anomalies in the weakly adiabatic regime of $\omega^{ad} \approx 1$.

## III. MEAN-FIELD THEORY OF SINGLE-CELL DYNAMICS

A transparent way to treat the master equation (3) is to use the analogy to quantum mechanics[19]. The difference operations in Eq.3 can be expressed by using the creation operator $a^\dagger$ and the annihilation operator $a$ with $[a, a^\dagger] = 1$. Such a notation was introduced by Doi[20] to describe the diffusion limited reactions and was used by Sasai and Wolynes[1] to describe gene switches. The state vector is introduced as $|\psi_\alpha(t)>=\Sigma_n P_\alpha(n,t)|n>$, where $|n>$ is defined by $a^\dagger\ |n>=|n+1>$ and $a|n>=n|n-1>$. $|\psi_\alpha(t)>$ has all the information of the system at time $t$ and can be used as a generating function for $P_\alpha(n,t)$[21]. We should note $<n|m>=\delta_{nm}n!$ and $<0|e^a|n>=1$, so that $|\psi_\alpha(t)>$ is normalized as $<0|e^a|\psi_\alpha(t)>=\Sigma_n P_\alpha(n,t)=1$.

By using $|\boldsymbol{\psi}(t)>=\ (|\psi_1(t)>,|\psi_0(t)>)^{\mathrm{t}}$ and a spinor "Hamiltonian" $\Omega$, Eq.3 is written as $(1/k)\partial|\boldsymbol{\psi}(t)>/\partial t=\Omega|\boldsymbol{\psi}(t)>$. The factor $1/k$ can be absorbed by scaling time as $kt\to t$, leading to $\partial|\boldsymbol{\psi}(t)>/\partial t=\Omega|\boldsymbol{\psi}(t)>$. $\Omega$ is a non-Hermitian operator,

$$\Omega=\begin{pmatrix} X^{\mathrm{ad}}+\delta X & 0 \\ 0 & X^{\mathrm{ad}}-\delta X \end{pmatrix}\left((a^+)^{N_b}-1\right)+a-a^+a+\omega^{\mathrm{ad}}\begin{pmatrix} -(a^+)^2a^2/X^{\mathrm{eq}} & (a^+)^2 \\ a^2/X^{\mathrm{eq}} & -1 \end{pmatrix}. \quad (4)$$

We introduce external fields, $J_\mu$ and $H_\mu$, and write $\overline{\Omega}=\Omega+\sum_\mu \hat{A}^\mu J_\mu(t)+\sum_\mu \hat{B}^\mu H_\mu(t)$, where $\hat{A}^\mu$ is the operator representation of $A^\mu$ in Eq.1 and $\hat{B}^\mu$ is the operator which is conjugate to $H_\mu$. For example, we can assign $\hat{A}^n=a^+a+1-\sigma_z$ with $\sigma_z=\begin{pmatrix}1 & 0\\ 0 & -1\end{pmatrix}$ for the total number of protein molecules, or $\hat{A}^\alpha=(1+\sigma_z)/2$ for the probability of the DNA state to be on. Various types of response functions can be defined by choosing appropriate $H_\mu$ in Eq.2. $g_\alpha\to g_\alpha+H_g$ and $k\to k+H_k$ are possible examples of perturbation. The conjugate quantities to them are $\hat{B}^g=\left((a^+)^{N_b}-1\right)/k$ and $\hat{B}^k=a-a^+a$, respectively.

We define the generating function, $Z=<0|e^a\hat{T}\exp(\int_{t_0}^{t_1}\overline{\Omega}dt)|\boldsymbol{\psi}(t_0)>$, where $\hat{T}$ expresses the operation of time-ordering product. Then, single or multiple-time expectation values can be calculated as $<\hat{A}^\mu(t)>=\delta\ln Z/\delta J_\mu(t)|_{J=H=0}$ and

$$C^{\mu\nu}(t,t')=<\hat{A}^\mu(t)\hat{A}^\nu(t')>-<\hat{A}^\mu(t)><\hat{A}^\nu(t')>=\delta^2\ln Z/\delta J_\mu(t)\delta J_\nu(t')|_{J=H=0},$$

$$R^{\mu\nu}(t,t')=<\hat{A}^\mu(t)\hat{B}^\nu(t')>-<\hat{A}^\mu(t)><\hat{B}^\nu(t')>=\partial^2\ln Z/\partial J_\mu(t)\partial H_\nu(t')|_{J=H=0},$$

(5)

for $t_1 > t > t' > t_0$. $Z$ can be written in a path-integral form as

$$Z = \text{const.}\int DcDc^* D\psi_1 D\psi_1^* D\psi_0 D\psi_0^* \exp\left(-\int_{t_0}^{t_1} dtL\right), \tag{6}$$

where $c$, $c^*$, $\psi_\alpha$, and $\psi_\alpha^*$ are independent functions of time, $DcDc^* = \lim_{N\to\infty}\prod_{r=1}^{N-1}(1+c^*(t_r)c(t_r))^{-2}dc(t_r)dc^*(t_r)$, $D\psi_\alpha D\psi_\alpha^* = \lim_{N\to\infty}\prod_{r=1}^{N-1}d\psi_\alpha(t_r)d\psi_\alpha^*(t_r)$, $t_r = t_0 + r\Delta t$, and $N\Delta t = t_1 - t_0$. The "Lagrangean" $L$ is

$$\begin{aligned}
L = \; & \frac{1}{1+cc^*}\left[\psi_1^*\left(\frac{d\psi_1}{dt}+\psi_1\right)-(X^{\text{ad}}+\delta X)\left\{(\psi_1^*+1)^{N_b}-1\right\}+\frac{\omega^{\text{ad}}}{X^{\text{eq}}}(\psi_1^*+1)^2\psi_1^{\,2}\right] \\
& +\frac{cc^*}{1+cc^*}\left[\psi_0^*\left(\frac{d\psi_0}{dt}+\psi_0\right)-(X^{\text{ad}}-\delta X)\left\{(\psi_0^*+1)^{N_b}-1\right\}+\omega^{\text{ad}}\right] \\
& -\frac{1}{1+cc^*}\sqrt{\frac{c}{c^*}}\frac{\omega^{\text{ad}}}{X^{\text{eq}}}\psi_1^{\,2}\exp\left[\psi_0^*(\psi_1-\psi_0)\right] \\
& -\frac{cc^*}{1+cc^*}\sqrt{\frac{c^*}{c}}\omega^{\text{ad}}(\psi_1^*+1)^2\exp\left[\psi_1^*(\psi_0-\psi_1)\right] \\
& -\frac{c^*}{1+cc^*}\frac{dc}{dt}+\sum_\mu \bar{A}^\mu J_\mu(t)+\sum_\mu \bar{B}^\mu H_\mu(t)\;,
\end{aligned} \tag{7}$$

where $\bar{A}^\mu$ and $\bar{B}^\mu$ are functions of $c$, $c^*$, $\psi_\alpha$, and $\psi_\alpha^*$. See APPENDIX for derivation of Eq.7 and the precise functional forms of $\bar{A}^\mu$ and $\bar{B}^\mu$.

The most weighted path in Eq.6 is obtained by taking variation, $\delta L/\delta c = \delta L/\delta c^* = \delta L/\delta\psi_\alpha = \delta L/\delta\psi_\alpha^* = 0$. When $J = H = 0$, the solution $c = c^* = c^{\text{cl}}$, $\psi_\alpha = X_\alpha$ and $\psi_\alpha^* = 0$ of the variational equation satisfy

$$D_1 \frac{dX_1}{dt} = \left\{ N_b (X^{\mathrm{ad}} + \delta X) - X_1 - 2\frac{\omega^{\mathrm{ad}}}{X^{\mathrm{eq}}} X_1{}^2 \right\} D_1 + \omega^{\mathrm{ad}} (2 + X_0 - X_1) D_0 ,$$

$$D_0 \frac{dX_0}{dt} = \left\{ N_b (X^{\mathrm{ad}} - \delta X) - X_0 \right\} D_0 + \frac{\omega^{\mathrm{ad}}}{X^{\mathrm{eq}}} X_1{}^2 (X_1 - X_0) D_1 ,$$

$$\frac{dD_1}{dt} = -\frac{\omega^{\mathrm{ad}}}{X^{\mathrm{eq}}} X_1{}^2 D_1 + \omega^{ad} D_0 , \tag{8}$$

where $D_1 = (1 + c^{\mathrm{cl}})^{-2}$ is the probability that the DNA state is on in this approximation and $D_0 = 1 - D_1$. $X_\alpha$ in Eq.8 can be interpreted as the expectation number of protein molecules when the DNA state is $\alpha$. Eq.8 has the same form as the equation obtained by Eyink's variational method under the coherent state ansatz[1] for the state vector $| \boldsymbol{\psi}(t) >$.

Dynamical fluctuations can be treated by collecting paths in Eq.6 around the most weighted path of Eq.8. In the adiabatic limit of large $\omega^{\mathrm{ad}}$ and large $X^{\mathrm{ad}}$ a good approximation should be obtained by considering fluctuations in $\psi_\alpha$ and $\psi_\alpha^*$ and neglecting fluctuations in $c$ and $c^*$. By writing $\psi_\alpha = X_\alpha + \hat{\psi}_\alpha$ and $\psi_\alpha^* = \hat{\psi}_\alpha^*$, and retaining the lowest order terms of $\hat{\psi}_\alpha$ and $\hat{\psi}_\alpha^*$ in $L$, the Langevin equation for the protein number fluctuation is derived from Eq.6. This treatment is equivalent to the adiabatic approximation often adopted in theoretical models[3,4].

Here, we do not take this approximation but take account of effects beyond this limit: In the weakly adiabatic regime of moderate $\omega^{\mathrm{ad}}$ or nonadiabatic regime of small $\omega^{\mathrm{ad}}$, fluctuations of $c$ and $c^*$ should be important. We incorporate those nonadiabatic fluctuations of the DNA state by expressing $c = c^{cl} + \hat{x} + \hat{y}$ and $c^* = c^{cl} + \hat{x} - \hat{y}$ to expand $L$ to the lowest order of $\hat{x}$, $\hat{y}$, $\hat{\psi}_\alpha$, and $\hat{\psi}_\alpha^*$. This is a Gaussian mean-field approximation for the path integral of Eq.6. This expansion should be reasonable when either $\omega^{\mathrm{ad}}$ or $X^{\mathrm{ad}}$ is not extremely small. As in many other mean-field theories in statistical physics, however, we may be able to expect that the approximation gives a qualitatively meaningful results even beyond the regime that the quantitative accuracy is assured. In the following, we show that the present mean-field treatment indeed gives semi-quantitative results even for the small $\omega^{\mathrm{ad}}$ or small $X^{\mathrm{ad}}$ regime and provides a guidance for further quantitative calculations of single-cell dynamics.

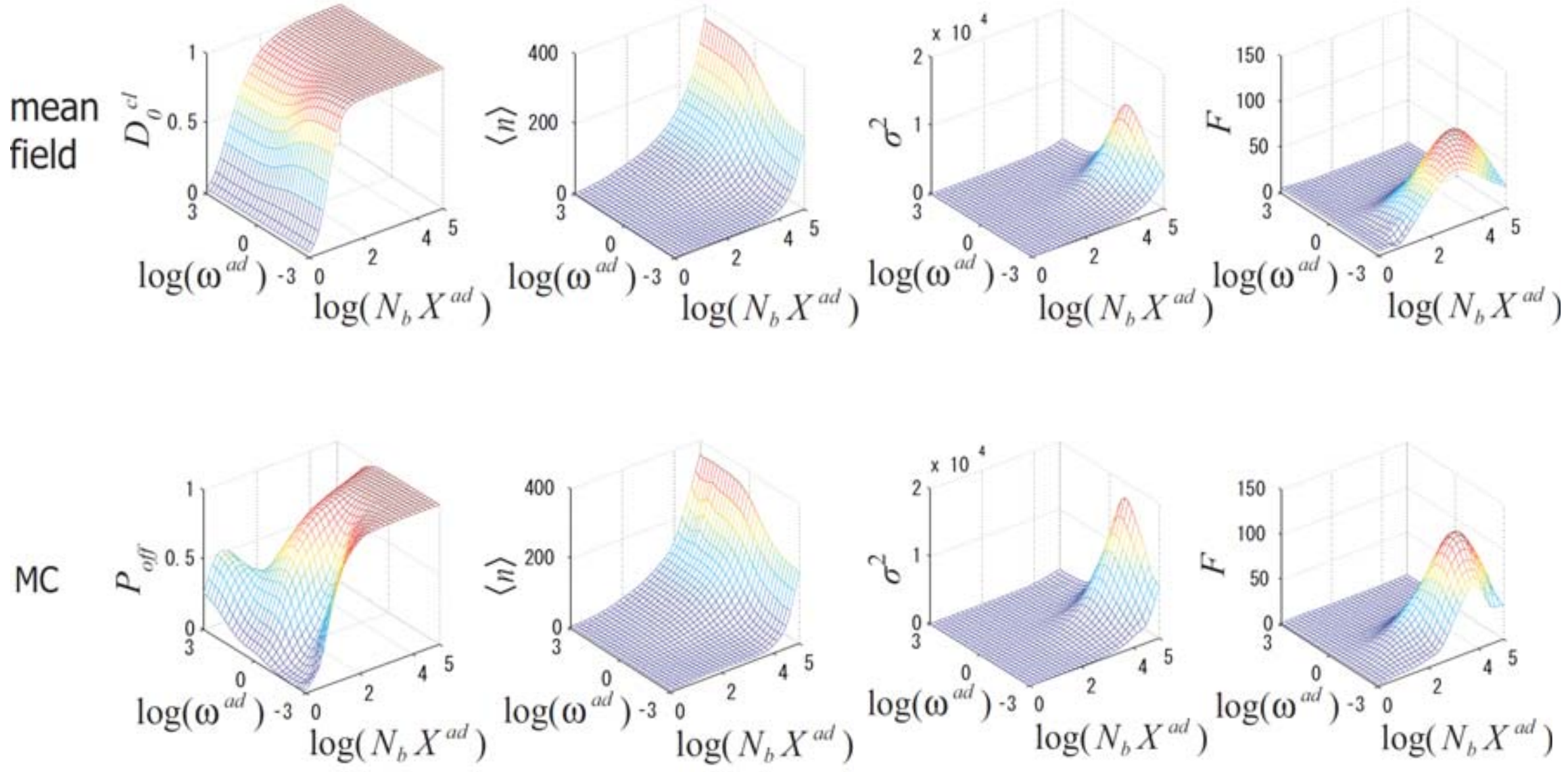

Figure 2

Dependence of static quantities and strength of fluctuationonadiabaticity, $\omega^{\rm ad}$, and the typical number of protein molecules, $N_bX^{\rm ad}$. From left to right, the probability of the DNA state to be off ($D_0$), the average ($<n>$) and variance ($\sigma^2$) of the protein number, and the strength of fluctuation ($F = \sigma^2/<n>$) are plotted on the plane of $\log_{10}\omega^{\rm ad}$ and $\log_{10}N_bX^{\rm ad}$. Results of the mean-field theory (upper figures) and those of the MC calculation (lower figures) are compared.

## IV. RESULTS

### A. Static quantities and strength of fluctuation

Before addressing problems in dynamics, we can assess the approximation used here by comparing it with the numerical Monte Carlo (MC) results obtained by applying the Gillespie algorithm[22] to Eq.3. In Fig.2 the probability for the DNA state to be off, $D_0$, and the average protein number, $<n>=D_0(X_0+2)+D_1X_1$ are calculated from the stationary solution of Eq.8, and are compared with the corresponding MC results. A qualitative agreement for $D_0$ and a good quantitative agreement for $<n>$ are found between two methods of calculation: As $X^{\rm ad}$ increases, $<n>$ increases as expected, which then suppresses the transcription to make $D_0$ large. $D_0 \approx 1$ when $N_bX^{\rm ad}$ is as large as $N_bX^{\rm ad} >> (X^{\rm eq})^{1/2} = 10^{-1}$. Even in this large $D_0$ regime, $<n>$ is kept large due to the bust production of protein in the infrequently occurring DNA on state.

Also compared in Fig.2 are variance in the protein number, $\sigma^2 = <n^2> - <n>^2$, and the strength of fluctuation, $F = \sigma^2 / <n>$. The results of $\sigma^2$ and $F$ calculated with the mean-field approximation in Eq.6 agree semi-quantitatively with the MC results:

When $\omega^{\rm ad}$ is large, $\sigma^2$ and $F$ are suppressed to be small due to the negative feedback action of the circuit. $\sigma^2$ is large, however, at the nonadiabatic regime of $\omega^{\rm ad} \approx 10^{-1}$ for large $X^{\rm ad}$, where the DNA state is kept off for long duration due to the small $\omega^{\rm ad}$, but protein is produced in bursts in the on state due to the large $X^{\rm ad}$, which leads to the large fluctuation in the protein number. For $\omega^{\rm ad} << 10^{-1}$ with large $X^{\rm ad}$, the burst expression occurs with much less frequency, leading to smaller $\sigma^2$. $F$ decreases for large enough $X^{\rm ad}$ due to the sharp increase in $<n>$, yielding a peak of $F$ as shown in Fig.2.

Thus, we found a distinct enhancement of noise caused by the slow DNA fluctuation, which is the *static nonadiabatic anomaly* discussed in Section II: The mean-field theory predicted that the strength of noise in the protein number begins to increase when the DNA fluctuation becomes as slow as the protein number degradation with $\omega^{\rm ad} < 1$ and the typical protein number is $N_b X^{\rm ad} > (X^{\rm eq})^{1/2}$. The strength of noise is largest when the DNA fluctuation is as slow as $\omega^{\rm ad} \approx 10^{-1}$ and the protein number is $N_b X^{\rm ad} \approx 10^4$. This anomaly in the noise strength is similar to what was found in the exact solution of the model of Ref.16. Instead of binding of dimer repressor with the rate $hn(n-1)$ in the present model, the model of Ref.16 was based on the assumption of binding of monomer repressor. Despite this difference, the apparent increase of $\sigma^2$ at the moderately small $\omega^{\rm ad}$ region of the weakly adiabatic regime is common to these two models and shows a symptom of *static nonadiabatic anomaly*.

**B. Correlation and response functions**

Single-cell dynamics can be analyzed through $C$ and $R$. When the system fluctuates around the stationary state under the condition of $t > t' >> t_0$, $C$ and $R$ should depend on $t-t'$ as $C(t-t')$ and $R(t-t')$. We here show the results for correlation of fluctuation in the protein number, $C^{nn}(t-t')$, by employing $\hat{A}^n = a^+ a + 1 - \sigma_z$, and response of the protein number to the perturbation of the protein degradation rate, $R^{nk}(t-t')$, by employing $\hat{A}^n = a^+ a + 1 - \sigma_z$ and $\hat{B}^k = a - a^+ a$ in Eq.5. In the mean-field results, both $C^{nn}(t-t')$ and $R^{nk}(t-t')$ have either of two functional forms, $Q_1 e^{-(t-t')/\tau_1} + Q_2 e^{-(t-t')/\tau_2} + Q_3 e^{-(t-t')/\tau_3}$ or $Q_1 e^{-(t-t')/\tau_1} + Q_2 \cos(\eta(t-t') - \lambda) \cdot e^{-(t-t')/\tau_2}$ .
The numerical MC results can also be fitted by either of these two functions.

In Fig.3, it is shown that $C$ and $R$ can be practically fitted by a single exponential

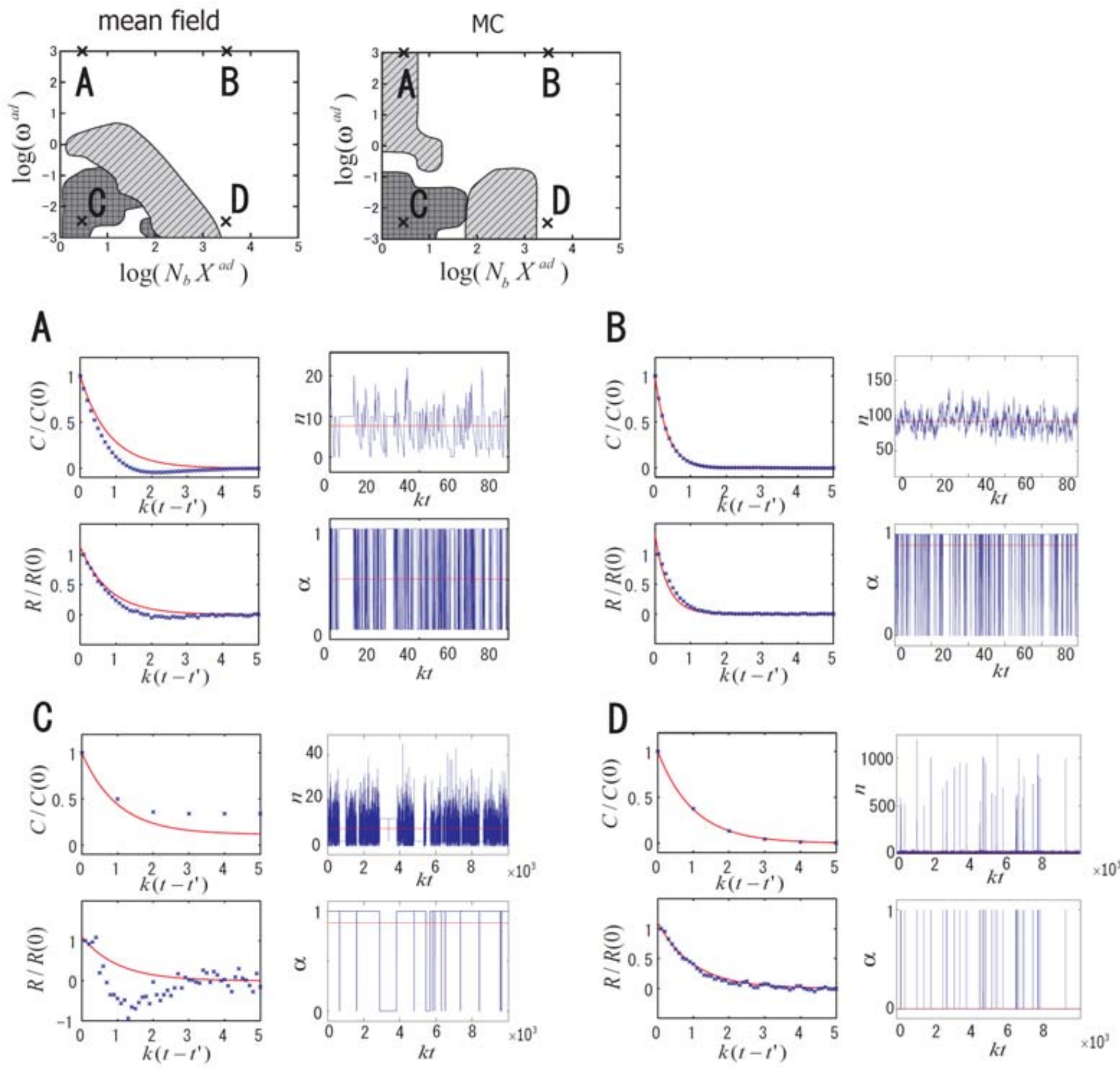

Figure 3

Functional form of $C$($t$-$t$') and examples of $C$($t$-$t$'), $R$($t$-$t$') and MC trajectories. In top two figures, functional form of $C$($t$-$t$') obtained from the mean-field theory and that of the MC calculation are shown on the plane of $\log_{10}\omega^{ad}$ and $\log_{10}N_bX^{ad}$. In the white region, $C$($t$-$t$') is dominated by a single exponential function and other minor components contribute less than 1% in $C$($t$-$t$'). In the region shaded with slanting lines, $C$($t$-$t$') has an oscillatory component and in the cross-hatched region, $C$($t$-$t$') is composed of multiple exponential functions. For the points A, B, C, and D designated on the plane, $C$($t$-$t$')/$C$(0) and $R$($t$-$t$')/$R$(0) calculated by the mean-field theory (red lines) and those of the MC results (crosses) are compared. Also exemplified are MC trajectories of the protein number, $n$, and the DNA state, $\alpha$, shown as functions of time, $k$($t$-$t$'). The horizontal red lines in the figures of $n$ and $\alpha$ denote their averaged values.

function in the wide parameter region because $|Q_1| >> |Q_2|$ and $|Q_1| >> |Q_3|$ there. In such region of single-exponential decay, the mean-field theory and the MC results show quantitatively good agreement for $C$ and $R$. When $\omega^{ad}$ and $X^{ad}$ are small, $|Q_1| \approx |Q_2|$ or $|Q_1| \approx |Q_3|$, so that $C$ and $R$ are fitted by multiple exponential or oscillating functions. In these regions of non-exponential decay, the functional form of $C$ or $R$ obtained from the mean-field theory does not necessarily reproduce the MC results as shown in Fig.3. This disagreement was expected because the mean-field treatment should not be safe for the case $\omega^{ad}$ or $X^{ad}$ is very small. Nevertheless, the mean-field calculation successfully predicts the existence of the region of this multi-exponential or oscillating decay, and gives a guideline for extensive numerical MC calculation.

Also shown in Fig.3 are examples of trajectories in the MC calculation. In the region of multi-exponential decay, the protein number fluctuation in the trajectory is intermittent, which should lead to the slow relaxation in $C$ and $R$. This slow relaxation is quantitatively confirmed by calculating relaxation times, $\tau_C$ obtained from $C$ and $\tau_R$ obtained from $R$. As shown in Fig.4, $\tau_C$ and $\tau_R$ grow very large in the region of small $\omega^{ad}$ and small $X^{ad}$ with $\omega^{ad} < 10^{-1}$ and $N_b X^{ad} < (X^{eq})^{1/2}$ in both the mean-field and MC results. In the large $\omega^{ad}$ and $X^{ad}$ region of the adiabatic regime, the rapid relaxation is realized due to the negative feedback action in the circuit[2], but in the small $\omega^{ad}$ and $X^{ad}$ region of the nonadiabatic regime, the slow DNA fluctuation manifests itself in the intermittent protein number fluctuation and yields *dynamical nonadiabatic anomaly*.

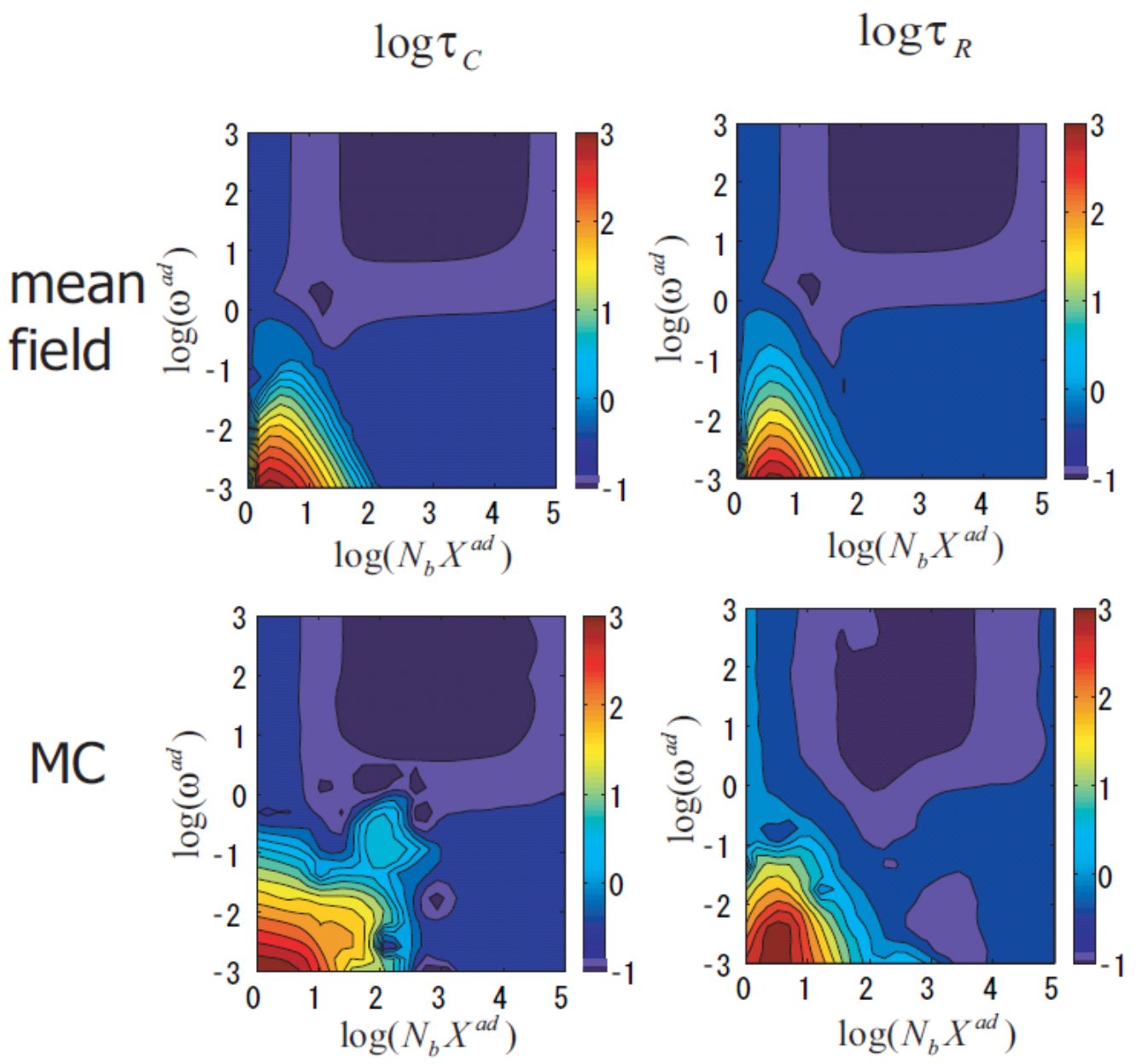

Figure 4
The relaxation time, $\tau_C$, calculated from $C(t\text{-}t')$ and the relaxation time, $\tau_R$, calculated from $R(t\text{-}t')$ are plotted with the unit of $1/k$ on the plane of $\log_{10}\omega^{ad}$ and $\log_{10}N_b X^{ad}$. The results of the mean-field theory (upper figures) and the MC results (lower figures) are compared.

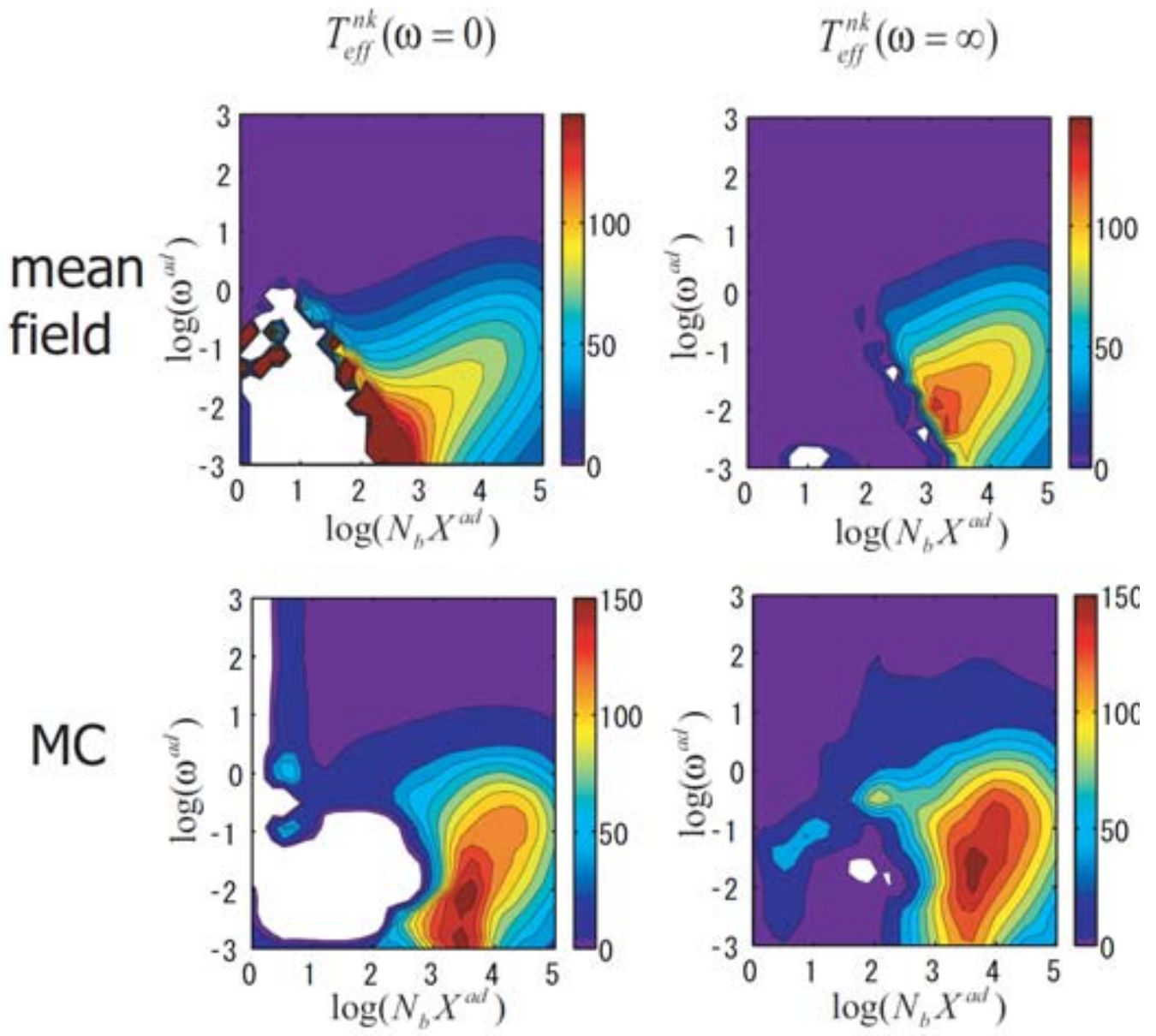

Figure 5

Effective temperature, $T_{\mathrm{eff}}^{nk}(\omega)$, defined by Eq.9 is plotted for $\omega = 0$ and $\omega = \infty$. $T_{\mathrm{eff}}^{nk}(\omega) < 0$ in the white region. The results of the mean-field theory (upper figures) and the MC results (lower figures) are compared.

**C. Effective temperature**

In systems near equilibrium, the fluctuation-response relation (FRR) holds as $R(t-t') = -(1/T)dC(t-t')/dt$, where $T$ is temperature in the unit of $k_B = 1$. In contrast, gene circuits are far from equilibrium and FFR is violated because the detailed balance is not fulfilled. In recent statistical mechanics of disordered systems[23], however, FRR is often extended to describe the far-from-equilibrium dynamics by introducing the effective temperature, $T_{\mathrm{eff}}$, as

$$\mathrm{Im}\,\tilde{R}^{\mu\nu}(\omega) = \frac{\omega}{2T_{\mathrm{eff}}^{\mu\nu}(\omega)}\tilde{C}^{\mu\mu}(\omega), \tag{9}$$

where $\tilde{R}^{\mu\nu}(\omega)$ and $\tilde{C}^{\mu\nu}(\omega)$ are Fourier transforms of $R^{\mu\mu}(t)$ and $C^{\mu\nu}(t)$, respectively. $T_{\mathrm{eff}}^{\mu\nu}(\omega)$ is a fictitious temperature introduced by Eq.9 and is generally a function of frequency $\omega$ and $\mu\nu$.

It is a challenging problem to investigate the possible physical meaning of thus defined effective temperature in gene circuits[24]. We here calculate $T_{\mathrm{eff}}^{nk}$ by using Eq.9 as a definition of $T_{\mathrm{eff}}^{nk}$ for the circuit of a self-regulating gene. In Fig.5 $T_{\mathrm{eff}}^{nk}(\omega = 0)$ and $T_{\rm eff}^{nk}(\omega=\infty)$ calculated with the mean-field and MC methods are shown. In the region of single-exponential decay, $T_{\rm eff}^{nk}$ has a peak similar to $F$ in Fig.2, showing the effective temperature obtained here designates the strength of fluctuation in the region of exponential decay and is usable to detect *static nonadiabatic anomaly*. In the region of small $\omega^{\rm ad}$ and $X^{\rm ad}$, on the other hand, $T_{\rm eff}^{nk}(\omega)$ shows a resonance of hyperbolic type with alternated sign around a certain frequency of $\omega=\omega_0$. Tails of this resonance are visible in $T_{\rm eff}^{nk}(\omega=0)$ for small $\omega^{\rm ad}$ and $X^{\rm ad}$, which designates a sign of *dynamical nonadiabatic anomaly* in this parameter region.

## V. SUMMARY AND DISCUSSION

In this paper we developed a nonadiabatic dynamical mean-field theory and applied it to a self-regulating gene circuit. The mean-field theory revealed rich effects of the DNA state fluctuation on single-cell dynamics including *static* and *dynamical nonadiabatic anomalies*. The rich nonadiabatic phenomena were also confirmed by the numerical MC calculation of the master equation of the model. We should note that these anomalies were not found with the conventional theories developed for the strong adiabatic limit. Thus, combined use of the analytic theory based on the path-integral representation of stochastic dynamics and the MC simulation is of great use to investigate static and dynamical anomalies in gene circuit. As the system size grows, however, the MC calculation requires much computational time and rapidly becomes difficult. The analytical mean-field method developed in this paper is applicable to more complex gene networks. Such analytical predictions should give a quick overview of single-cell dynamics over wide ranges of parameters and provide a framework to perform further extensive numerical calculations.

It is left for further study to investigate how the evolutionarily designed gene circuit avoids or utilizes *static* or *dynamical nonadiabatic anomalies*. It would be interesting to examine the possibility of whether the rate of the DNA state fluctuation or that of the protein number change is controlled in cells[25] to adapt to the environment through the mechanism of nonadiabatic anomalies. More experimental data on the single-cell dynamics will be obtained, which should increase our knowledge of dynamical heterogeneity, and emphasize the need for mathematical theory based on the ideas and methods of statistical many-body physics.

**APPENDIX**

In this appendix we derive Eq.7 of Section III. The generating function is written as

$$Z =< 0 | \hat{T} \exp(\int \tilde{\Omega} dt) | \tilde{\boldsymbol{\psi}} > = \lim_{N\to\infty} < 0 | \prod_{r=1}^{N-1} (1+\tilde{\Omega}(r)\Delta t) | \tilde{\boldsymbol{\psi}} >, \qquad (10)$$

where $\tilde{\Omega} = e^a \overline{\Omega} e^{-a}$ is a two-dimensional spinor operator, $| \tilde{\boldsymbol{\psi}} >= e^a | \boldsymbol{\psi} >$, and $N\Delta t$ = $t_1 - t_0$ with $t_1 > t > t' > t_0$. We use identity relations,

$$(1/\pi)\int d\psi_\alpha d\psi_\alpha^* \, |\psi'_\alpha >< \overline{\psi}_\alpha | = 1,$$

$$(1/2\pi)\int_0^\pi \sin\theta d\theta \int_0^{2\pi} d\varphi \, | s' >< \overline{s} | = 1, \qquad (11)$$

where $|\psi'_\alpha >= \exp(a^+ \psi_\alpha) | 0 >$ , $< \overline{\psi}_\alpha | =< 0 | \exp(a\psi_\alpha^* - \psi_\alpha \psi_\alpha^*)$ , $| s' >= \left(e^{i\varphi/2}\cos^2\theta/2, \;\; e^{-i\varphi/2}\sin^2\theta/2\right)^{\mathrm{t}}$, and $< \overline{s} | = \left(e^{-i\varphi/2}, \;\; e^{i\varphi/2}\right)$. Notice that the integrand of Eq.11 is not Hermitian which accommodate the non-Hermitian form of the operator $\tilde{\Omega}$. Inserting Eq.11 into Eq.10 and taking the limit of $N \to \infty$, Z can be written in a path-integral form,

$$Z = K(N)K(0)\int D\theta D\varphi D\psi_1 D\psi_1^* D\psi_0 D\psi_0^* \exp\left(-\int_{t_0}^{t} dtL\right), \qquad (12)$$

where $D\theta = \lim_{N\to\infty}\prod_{r=1}^{N-1} (1/2)\sin\theta(r) d\theta(r)$, $D\varphi = \lim_{N\to\infty}\prod_{r=1}^{N-1} (1/\pi) d\varphi(r)$, etc. and $K(N) = e^{i\varphi(N)/2}\cos^2(\theta(N)/2) + e^{-i\varphi(N)/2}\sin^2(\theta(N)/2)$ and $K(0) = e^{-i\varphi(1)/2}\langle \overline{\psi}_1(1) | \tilde{\psi}_1(0) \rangle + e^{i\varphi(1)/2}\langle \overline{\psi}_0(1) | \tilde{\psi}_0(0) \rangle$. When $J_\mu = H_\mu = 0$, "Lagrangean" $L$ is $L = L_0$ with

$$L_0 = \frac{1+\cos\theta}{2}\left[\psi_1^*\left(\frac{d\psi_1}{dt} + \psi_1\right) - (X^{\mathrm{ad}} + \delta X)\left\{(\psi_1^* + 1)^{N_b} - 1\right\} + \frac{\omega^{\mathrm{ad}}}{X^{\mathrm{eq}}}(\psi_1^* + 1)^2 \psi_1{}^2\right]$$

$$+\frac{1-\cos\theta}{2}\left[\psi_0^*\left(\frac{d\psi_0}{dt}+\psi_0\right)-(X^{\mathrm{ad}}-\delta X)\left\{(\psi_0^*+1)^{N_b}-1\right\}+\omega^{\mathrm{ad}}\right]$$

$$-\frac{1+\cos\theta}{2}\frac{\omega^{\mathrm{ad}}}{X^{\mathrm{eq}}}\psi_1{}^2\exp\left[\psi_0^*(\psi_1-\psi_0)\right]\exp(i\varphi)$$

$$-\frac{1-\cos\theta}{2}\omega^{\mathrm{ad}}(\psi_1^*+1)^2\exp\left[\psi_1^*(\psi_0-\psi_1)\right]\exp(-i\varphi)$$

$$+\frac{i}{2}\frac{d\varphi}{dt}\cos\theta. \qquad (13)$$

$\theta$ and $\varphi$ represent direction of a spin vector. $L$ involves imaginary values, which may make the physical interpretation of this "Lagrangean" unclear. We thus transform $\theta$ and $\varphi$ to $c=\tan(\theta/2)\exp(i\varphi)$ and $c^*=\tan(\theta/2)\exp(i\varphi)$. Then, $(1/2\pi)\sin\theta d\theta d\varphi=(i/\pi)(1+c^*c)^{-2}dcdc^*$ The Lagrangean (13) is

$$L_0 = \frac{1}{1+cc^*}\left[\psi_1^*\left(\frac{d\psi_1}{dt}+\psi_1\right)-(X^{\mathrm{ad}}+\delta X)\left\{(\psi_1^*+1)^{N_b}-1\right\}+\frac{\omega^{\mathrm{ad}}}{X^{\mathrm{eq}}}(\psi_1^*+1)^2\psi_1{}^2\right]$$

$$+\frac{cc^*}{1+cc^*}\left[\psi_0^*\left(\frac{d\psi_0}{dt}+\psi_0\right)-(X^{\mathrm{ad}}-\delta X)\left\{(\psi_0^*+1)^{N_b}-1\right\}+\omega^{\mathrm{ad}}\right]$$

$$-\frac{1}{1+cc^*}\sqrt{\frac{c}{c^*}}\frac{\omega^{\mathrm{ad}}}{X^{\mathrm{eq}}}\psi_1{}^2\exp\left[\psi_0^*(\psi_1-\psi_0)\right]$$

$$-\frac{cc^*}{1+cc^*}\sqrt{\frac{c^*}{c}}\omega^{\mathrm{ad}}(\psi_1^*+1)^2\exp\left[\psi_1^*(\psi_0-\psi_1)\right]$$

$$-\frac{c^*}{1+cc^*}\frac{dc}{dt}, \qquad (14)$$

Notice that $c$ and $c^*$ or $\psi_\alpha$ and $\psi_\alpha^*$ are originally defined as pairs of complex conjugate. They can be interpreted, however, as independent real variables by analytically changing the routes of integration.

When $J_\mu \neq 0$ and $H_\mu \neq 0$, we should write $L=L_0+\sum_\mu \bar{A}^\mu J_\mu+\sum_\mu \bar{B}^\mu H_\mu$. If we use, $\hat{A}^n = a^+ a + 1 - \sigma_z$ with $\sigma_z = \begin{pmatrix} 1 & 0 \\ 0 & -1 \end{pmatrix}$ , $\hat{A}^\alpha = (1+\sigma_z)/2$ , $\hat{B}^g = \left((a^+)^{N_b} - 1\right)/k$ , and $\hat{B}^k = a - a^+ a$ , then Eq.14 becomes

$$L = L_0 + \overline{A}^n J_n + \overline{A}^\alpha J_\alpha + \overline{B}^g H_g + \overline{B}^k H_k , \qquad (15)$$

with $\overline{A}^n = \frac{1}{1+cc^*}\left[\psi_1^* \psi_1 + \psi_1\right] + \frac{cc^*}{1+cc^*}\left[\psi_0^* \psi_0 + \psi_0 + 2\right]$ , $\overline{A}^\alpha = \frac{1}{1+cc^*}$ , $\overline{B}^g =$ $\frac{1}{1+cc^*}\left\{\left(\psi_1^* + 1\right)^{N_b} - 1\right\} + \frac{cc^*}{1+cc^*}\left\{\left(\psi_0^* + 1\right)^{N_b} - 1\right\}$, and $\overline{B}^k = \frac{1}{1+cc^*}\psi_1^* \psi_1 + \frac{cc^*}{1+cc^*}\psi_0^* \psi_0$ .
Eq.15 is Eq. 7 in Section III

**Acknowledgement**

This work was supported by grants from the Ministry of Education, Culture, Sports, Science, and Technology, Japan, and by grants for the 21st century COE program for Frontiers of Computational Science. Y.O. thanks the Research Fellowships for Young Scientists from the Japan Society for the Promotion of Science.